# Flying on a Rainbow – A Solar-Driven Diffractive Sailcraft (Invited TVIW Paper)

Grover A. Swartzlander, Jr.

*Chester F. Carlson Center for Imaging Science*
*Rochester Institute of Technology, Rochester, NY 14623*
Corresponding author: grover.swartzlander@gmail.com

Abstract: Radiation pressure afforded by natural broadband sunlight upon a transmissive diffractive sail is theoretically and numerically investigated. A grating period of one micrometer is found to convert 83% of the solar black body spectrum into sailcraft momentum. Non-optimized orbit-raising trajectories for diffractive and reflective sails are compared. Potential advantages of diffractive sails are also described.

Recent and upcoming sailcraft demonstration missions are beginning to utilize the free and abundant momentum of solar photons for in-space navigation and propulsion [1-5]. An alternative to the longstanding assumption of a reflective sail [6,7] has recently been proposed [8], whereby a thin single order diffraction grating replaces the metal-coated polymer film. The net force on a single order diffraction grating owing to the broadband solar spectrum is reported here for a transmissive grating that is suitable for orbit-raising. Combining the grating equation and the solar black body spectral exitance, pressures that are comparable to a reflective sails are derived, but with different forcing laws. Solar-driven sails have been proposed for missions ranging from the near-Earth to the interstellar realms (see for example [9]). A simple example is reported here whereby the sail makes a synchronous transfer orbit between Earth and Mars. Such transfers may occur without changing the relative attitude of the sail with respect to the sunline if the force in the orbital direction is greater than the force in the sunline direction [8]. While both diffractive and reflective sails may satisfy this condition with solar illumination, the former brings additional advantages.

A beam of light with wavelength $\lambda$ is diffracted from a grating having a period $\Lambda$ according to a phase matching condition at the grating boundary, as dictated by Maxwell's equations. Assuming the grating momentum vector $\vec{K}$ is tangential to the surface, with magnitude $K = 2\pi/\Lambda$, the boundary condition may be expressed $(\vec{k}_i + m\vec{K}) \cdot \hat{p} = \vec{k}_m \cdot \hat{p}$ where $\vec{k}_i/k_0 = -\sin\theta_i \hat{p} + \cos\theta_i \hat{n}$ and $\vec{k}_m/k_0 = \sin\theta_m \hat{p} - \cos\theta_m \hat{n}$ are the incident and $m^{th}$ order diffracted wave vectors, respectively, $k_0 = 2\pi/\lambda$, and $\hat{p}$ and $\hat{n}$ are the respective parallel and normal unit vectors of the grating surface (see Fig. 1). The diffraction angle $\theta_m$ may be expressed in terms of sine and cosine functions:

$$\sin\theta_m = m\lambda/\Lambda - \sin\theta_i; \quad \cos\theta_m = \pm\left(1 - \sin^2\theta_m\right)^{1/2} \tag{1}$$

where the plus (minus) sign corresponds to a reflected (transmitted) grating order.

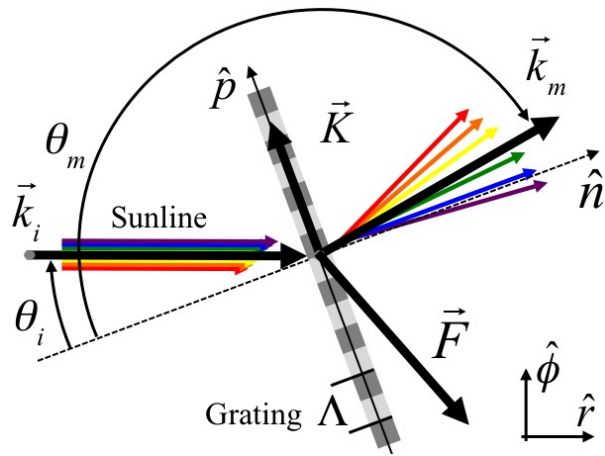

Fig. 1. Light incident upon a diffraction grating of period $\Lambda$, with the sunline $\hat{x}$ and surface normal $\hat{n}$ subtending the angle $\theta_i$. Incident, diffracted, grating wave vectors: $\vec{k}_i$, $\vec{k}_m$, $\vec{K}$. Surface unit vector $\hat{p}$. Diffraction angle, $\theta_m$. Radiation pressure force imparted to the grating, $\vec{F}$.

For a given wavelength the radiation pressure force on a grating may be expressed

$$\vec{F}_\lambda = \frac{IA\cos\theta_i}{c} \sum_m \eta_m \left( C_{i,m} \hat{n} - S_{i,m} \hat{p} \right) \qquad (2)$$

where $C_{i,m} = \cos\theta_i + \cos\theta_m$, $S_{i,m} = \sin\theta_i + \sin\theta_m = m\lambda/\Lambda$, $\eta_m$ is the fraction of incident beam power that is diffracted into the $m^{th}$ order, $I$ is the irradiance, $A$ is the area of the grating, and $c$ is the speed of light. For orbital dynamical modeling it is convenient to represent the force with respect to the optical axis (i.e., the sunline) coordinate system, where $\hat{r} = \sin\theta_i \hat{n} + \cos\theta_i \hat{p}$ and $\hat{\phi} = \cos\theta_i \hat{n} - \sin\theta_i \hat{p}$. Furthermore if the grating is exposed to a beam having a spectral irradiance distribution $B(\lambda)$, then the net force may be determined by integration:

$$F_\phi = \frac{A\cos\theta_i}{c} \sum_m \int B(\lambda)\eta_m(\lambda)\left(C_{i,m}\sin\theta_i - S_{i,m}\cos\theta_i\right)d\lambda \qquad (3a)$$

$$F_r = \frac{A\cos\theta_i}{c} \sum_m \int B(\lambda)\eta_m(\lambda)\left(C_{i,m}\cos\theta_i + S_{i,m}\sin\theta_i\right)d\lambda \qquad (3b)$$

where $\eta_m(\lambda)$ is the wavelength-dependent $m^{th}$ order diffraction efficiency. If $M(\lambda)$ represents the black body spectral exitance from the sun, then $B(\lambda) = (R_{sun}/r)^2 M(\lambda)$ where $R_{sun} = 6.957 \times 10^8 [m]$ is the solar radius, r is the distance between the sun and the sailcraft, and $M(\lambda) = (2\pi hc^2/\lambda^5)/(\exp(hc/\lambda k_B T)-1)$, where $h$ and $k_B$ are respectively the Planck and Boltzmann constants, and $T$=5778K is the effective temperature of the sun. The solar spectral irradiance at a distance of $r = 1$ [AU] ($1.496 \times 10^{11}[m]$) is shown in Fig. 2. Although the sun is not a perfect black body radiator, integration of $B(\lambda)$ across all wavelengths at 1 AU agrees well with the measured value of the "solar constant", $I_s = 1.37$ kW/m².

The allowed diffraction angles, dictated by Eq. (1), provide lower and upper wavelength bounds: $\lambda_{min} = 0$ and $\lambda_{max} = (\Lambda/m)(1+\sin\theta_i)$. Examples of the spectral diffraction range are evident in Fig. 2 for a grating period of $\Lambda$ = 1 µm, with a diffraction cut-off wavelength at $\lambda_{max} = 1.34$ µm. This period is an optimized value, described below, that makes use of roughly 83% of the available solar power. In comparison reflective sails may use a similar fraction after accounting for absorption, re-radiation, and the trade-off between the metallic layer thickness and its mass.

In practice modern gratings may be designed with meta-material engineering approaches to optimize the diffraction efficiency across a band of wavelengths. An ideal grating having 100% diffraction efficiency into the $m$=1 order is assumed below for all wavelengths up to $\lambda_{max}$. The solar radiation pressure force on a grating was numerically computed for the solar black body spectrum. Optimizing for both a large magnitude of transverse force and a small value of longitudinal force, $\Lambda$=1 [µm] and $\theta_i$ = 20° was determined, along with the corresponding radiation pressure spectral density distributions shown in Fig. 2. Integrating across wavelengths provides the values of the net force components: $F_\phi = -2.26 \times 10^{-6} [N/m^2]A$ and $F_r = 1.03 \times 10^{-6} [N/m^2]A$ at $r = 1$ AU. The corresponding force efficiencies may be expressed $\tilde{\eta}_\phi = F_\phi c/I_s A = -0.50$ and $\tilde{\eta}_r = F_r c/I_s A = 0.22$. This is a desired condition for orbit-raising (or lowering), i.e. $|\tilde{\eta}_\phi| > \tilde{\eta}_r$.

In comparison, the force efficiencies of an ideal non-absorbing, perfectly reflecting sail are given by $\tilde{\eta}'_r = 2\cos^3(\theta_i)$ and $\tilde{\eta}'_\phi = 2\cos^2(\theta_i)\sin(\theta_i)$ [9]. In this case the incident angle must exceed 45° to achieve the criterion $|\tilde{\eta}'_\phi/\tilde{\eta}'_r| > 1$. This does not make efficient use of the sail because the available driving power on the sail falls as $\cos(\theta_i)$. Nevertheless, the reflective sail achieves a transverse efficiency $\tilde{\eta}'_\phi = 0.50$ when $\theta_i$ = 56.9° with a corresponding longitudinal efficiency $\tilde{\eta}'_r = 0.326$. The latter value is larger and therefore less desirable than the corresponding value for the diffractive sail discussed above.

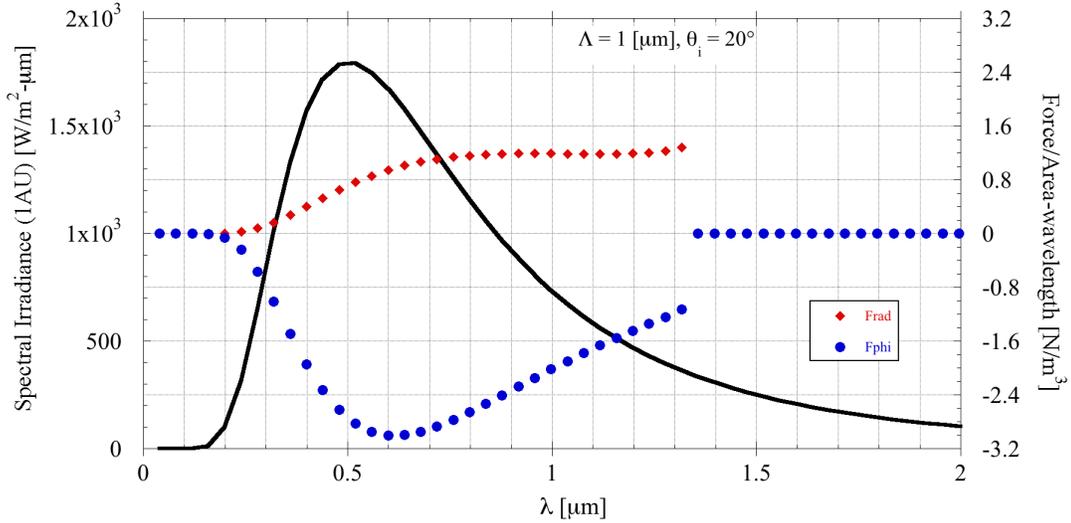

Fig. 2. Black body solar spectral irradiance at T=5778K (black line). Transverse (circles) and longitudinal (diamonds) solar radiation pressure spectral density distribution for an ideal single order grating of period Λ=1 μm and incidence angle $\theta_i$=20°.

The net force on the sailcraft from solar gravity and radiation pressure at any distance from the sun, $r$, may be expressed [9]

$$\vec{F} = \vec{F}_G + \vec{F}_{RP} = -G\tilde{m}Mr^{-2}\hat{r} + A(I_s/c)(R_E/r)^2\vec{\eta} = \tilde{m}\vec{a} \qquad (4)$$

where $R_E$ = 1 [AU], G is the universal gravitational constant, and $\tilde{m}$ and $M$ are the respective sailcraft and solar masses. The equation of motion may be expressed in normalized form:

$$\vec{F} = -\tilde{m}a_M(R_E/r)^2\left((1-\eta_r\sigma^*/2)\hat{r} - (\eta_\phi\sigma^*/2)\hat{\phi}\right) \qquad (5)$$

where $\sigma = \tilde{m}/A$ is the area density of the sailcraft, $\sigma^* = \sigma_{cr}/\sigma$ is the so-called lightness number, and $\sigma_{cr} = 2R_E^2 I_E/GMc = 1.54$ [g/m²] is a characteristic mass density, and $a_M = GM/R_E^2$. Note that as the areal density decreases, the lightness increases. A synchronous transfer orbit matches the orbital boundary conditions between planet A and planet B: $r(t=0) = R_A$, $\vec{v}(t=0) = v_A\hat{\phi}$, $r(t=T_0) = R_B$, $\vec{v}(t=T_0) = v_B\hat{\phi}$, where $v_{A,B} = \sqrt{GM/R_{A,B}}$ where $T_0$ is the transit time determined from numerical integration of Eq. (5). The sail is jettisoned or stowed at $T_0$ to maintain the desired orbit. Assuming circular co-planar orbits, synchronous or quasi-synchronous transfer orbits may be achieved from Earth to Mars if $\eta_\phi > \eta_r$ and $\eta_\phi\sigma^* \approx 0.0635$ [8]. To spiral outward the grating order is changed to m=-1, and the attitude is set to $\theta_i = -20°$ with respect to the sunline, resulting in a transverse (radial) force efficiency value of $\eta_\phi = +0.5$. ($\eta_r = +0.22$). The transfer requires a sailcraft areal density of roughly 12.1 [g/m²] (i.e, $\sigma^* = 0.127$). Using the fourth-order Runge-Kutta numerical integration technique [10], a value of $T_0$ =1.439 years (525.6 days) is found for an Earth to Mar transfer. The transfer time was similarly calculated to be four days longer for a reflective sail also having $\eta_\phi = +0.5$.

In principle any mission trajectory must allow adjustments of the force components to account for disturbances or to optimize the transfer time. Such a detailed study is beyond the scope of this report. Like a reflective sail, the force on a diffractive sail may be varied by changing the attitude of the sail with respect to the sunline. In this case the forcing law of a sun-lit grating of period Λ=1 μm is shown in Fig. 3. Clearly the first diffraction order provides the greatest magnitude of transverse force. What is more, by varying the attitude, say between -20° and 20°, the transverse force may be changed by roughly 50%. Note that negative diffraction orders produce the same lines, but with $\theta_i \to -\theta_i$.

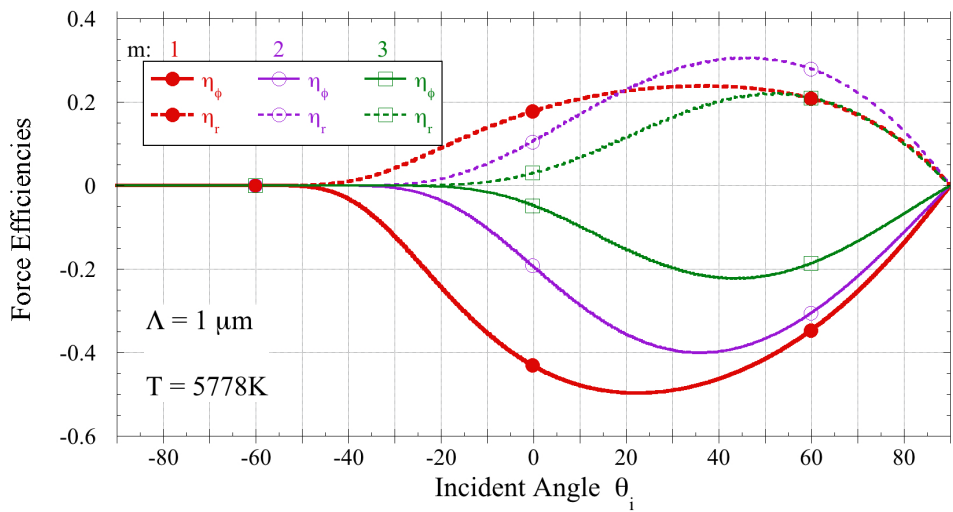

**Fig. 3**. Net transverse (solid lines) and longitudinal (dashed lines) force efficiencies of a solar-driven diffractive sail having a grating period Λ=1 μm for diffraction orders m=1,2,3 (corresponding to red, purple, and green lines).

One of the potential advantages of diffractive sails is the opportunity to change the diffractive properties by electro-optic means. This may allow a forcing law based on photonics rather than cumbersome mechanical control systems. Two different approaches to this solution may be described with the aid of Fig. 4, which plots the force efficiencies for the *m*=1 and -1 orders of a solar driven sail having an attitude of 20°. By actively varying the grating period, components of force may be directly controlled. In particular, the transverse force efficiency is found to vary linearly when the grating period varies between Λ = 1 and 2 μm (see straight dashed line in Fig. 4). Another proposed force control scheme involves switching the grating order from +1 to the -1. Changing the sign of the order changes the direction of the transverse force, thereby allowing navigation of the sailcraft.

Another advantage of a diffractive sailcraft is uncovered by the fact that the sail transmits photons which may be re-used to enhance the radiation pressure force or to generate photovoltaic energy for the communication and control bus. What is more, absorption within the diffractive sail may be negligible, thereby obviating heat stress and re-radiation problems of metallic films [9, 11]. Unfurling a diffractive sail in close proximity to the sun may therefore afford a route to interstellar space [12,13].

In summary the premise that sun-driven diffractive sails may replace reflective sails to achieve mission objectives such as orbit raising has been theoretically validated. The same argument may be made for orbit lowering (e.g., an Earth to Venus trajectory). A grating period of be Λ ≈ 1 μm was found to achieve a large transverse efficiency across the solar spectrum. The principle of non-mechanical navigation methods were proposed whereby either the grating period or grating order is varied, e.g., electro-optic means. Photonic and meta-materials research is needed to develop the required diffractive films that provide a high efficiency single diffraction order across the visible and near-infrared region of the solar spectrum. The film must also be able to withstand the harsh ultraviolet environment of outer space.

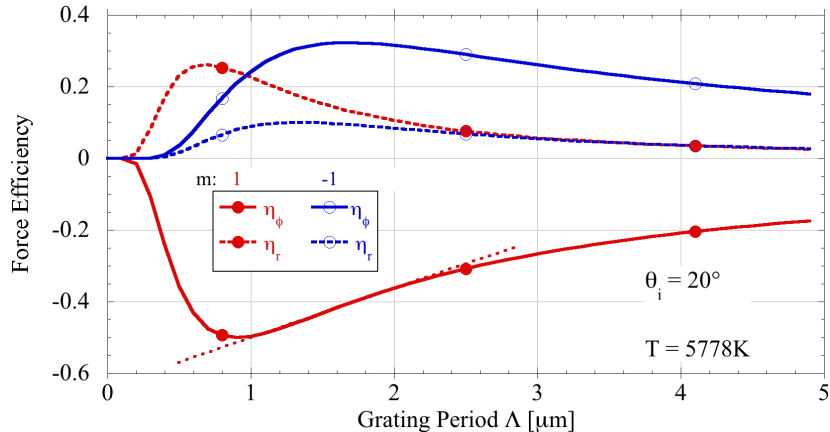

**Fig. 4**. Net transverse (solid lines) and longitudinal (dashed lines) force efficiencies of a solar-driven diffractive sail having a variable grating period for diffraction orders m=1 and -1 (respectively corresponding to red and blue lines).